\documentclass[twocolumn,preprintnumbers,amsmath,amssymb]{revtex4}
\usepackage{graphicx}
\usepackage{amssymb}

\usepackage{multirow}

\begin{document}

\title{Solving the encoding bottleneck: of the HHL algorithm, by the HHL
algorithm}
\author{Guang Ping He}
\email{hegp@mail.sysu.edu.cn}
\affiliation{School of Physics, Sun Yat-sen University, Guangzhou 510275, China}

\begin{abstract}
The Harrow-Hassidim-Lloyd (HHL) algorithm offers exponential speedup for
solving the quantum linear-system problem. But some caveats for the speedup
could be hard to met. One of the difficulties is the encoding bottleneck,
i.e., the efficient preparation of the initial quantum state. To prepare an
arbitrary $N$-dimensional state exactly, existing state-preparation
approaches generally require a runtime of $O(N)$, which will ruin the
speedup of the HHL algorithm. Here we show that the states can be prepared
approximately with a runtime of $O(poly(\log N))$ by employing a slightly
modified version of the HHL algorithm itself. Thus, applying this approach
to prepare the initial state of the original HHL algorithm can preserve the
exponential speedup advantage. It can also serve as a standalone solution
for other applications demanding fast state preparation.
\end{abstract}

\maketitle



\section{Introduction}

The success of Shor's factoring algorithm \cite{qi696} and Grover's search
algorithm \cite{qi2470} stimulated the rapid development of quantum
computation technology that lasted to this day. But as we entered the 21st
century, new quantum algorithms with comparable impacts have been rare to find.
The Harrow-Hassidim-Lloyd (HHL) algorithm \cite{qi2066} is one of the few
remarkable examples. It provides an elegant solution to the quantum version
of the linear-system problem (LSP), defined as \cite{qi2177}:

\textbf{QLSP}: Given a Hermitian $N\times N$ matrix $A$ and an $N$%
-dimensional unit vector $b=(b_{1},...,b_{N})^{T}$, find an $n$-qubit state $%
\left\vert \tilde{x}\right\rangle $ such that $\left\Vert \left\vert
x\right\rangle -\left\vert \tilde{x}\right\rangle \right\Vert \leq \epsilon $%
\ and%
\begin{equation}
Ax=b,  \label{HHL}
\end{equation}%
where $\epsilon $\ is the error and $N=2^{n}$.

The HHL algorithm can solve this problem in $O(poly(\log (N)))$ runtime,
while the best known classical algorithm for the classical LSP takes $%
O(poly(N))$ runtime. As LSP is one of the most basic problems in all of
science, the HHL algorithm is highly favored in the study of machine
learning \cite{qi1519,qi2462}, quantum walk \cite{qi2458}, computational
fluid dynamics \cite{qi2461}, quantum many body problem \cite{qi2463}, and
electromagnetic scattering cross-sections \cite{Ref5ofML391}, et al.

However, as pointed out in \cite{ml391}, this exponential speedup requires
the following caveats:

(i) The vector $b$ needs to be converted quickly into a quantum computer as
the $n$-qubit quantum state%
\begin{equation}
\left\vert b\right\rangle
=(b_{1},...,b_{N})^{T}=\sum\limits_{j=0}^{2^{n}-1}b_{j}\left\vert
j\right\rangle .
\end{equation}

(ii) The quantum computer needs to be able to apply unitary transformations
of the form $e^{-iAt}$ for various values of $t$.

(iii) The matrix $A$ needs to be \textquotedblleft
well-conditioned\textquotedblright , i.e., the condition number $\kappa $
(defined as the ratio in magnitude between the largest and smallest
eigenvalues of $A$) should not grow like $N^{c}$.

(iv) The quantum computer user can use the resultant quantum state $%
\left\vert x\right\rangle $\ directly, without needing to learn the value of
any specific entry of $x$.

Here we focus on the caveat (i). To initialize the quantum computer to the
state $\left\vert b\right\rangle $ corresponding to an arbitrary known
vector $b$ exactly, existing state-preparation approaches generally require $%
O(N)$ runtime, except for some special states \cite{qi2448,ml81,ml230,ml273}%
. Without a better solution, the exponential speedup of the HHL algorithm
will be ruined. This is called \textquotedblleft the encoding
bottleneck\textquotedblright ,
which seriously limits the application range of the HHL algorithm.

In this research, we will show that the state $\left\vert b\right\rangle $
can be prepared approximately to any desired nonzero error, with the runtime
speeded up to $O(poly(\log (N)))$. The approach is to apply the HHL
algorithm itself to prepare the initial state, with two minor modifications
only.

Moreover, the application of our approach is not limited to preparing the
initial state of the HHL algorithm. Any other task in need of the
preparation of quantum states can be benefited too, e.g., training quantum
machine learning models \cite{ml128,ml368,ml157,ml381,ml380,ml52,ml53} using
the amplitude encoding method \cite{ml118,ml269} for classical input data.

\section{Main idea}

Eq. (\ref{HHL}) means that the HHL algorithm is capable of finding an
approximation of the quantum state $\left\vert x\right\rangle $\ where $x$\
satisfies%
\begin{equation}
x=A^{-1}b,  \label{HHL-1}
\end{equation}%
given that $A$ and $b$ are known. Our goal is to find an algorithm to
prepare the state $\left\vert b\right\rangle $ efficiently.

While preparing $\left\vert b\right\rangle $\ for arbitrary values of $b$ is
hard in general, some special states does not have to take $O(poly(N))$
runtime to prepare. Therefore, we can start from such a state, then find an
effective method to turn it into any arbitrary $\left\vert b\right\rangle $.
Interestingly, with some minor twists, the HHL algorithm itself provides
such a solution, as elaborated below.

Since the classical entries $b_{1},...,b_{N}$ of the vector $b$\ is known,
let us define a matrix%
\begin{equation}
B=\left[
\begin{array}{ccccc}
b_{1} & 0 & \cdots & 0 & 0 \\
0 & b_{2} & \cdots & 0 & 0 \\
\vdots & \vdots & \ddots & \vdots & \vdots \\
0 & 0 & \cdots & b_{N-1} & 0 \\
0 & 0 & \cdots & 0 & b_{N}%
\end{array}%
\right]  \label{B}
\end{equation}%
and an $N$-dimensional vector%
\begin{equation}
h=(1,1,...,1,1)^{T}.
\end{equation}%
Then we have%
\begin{equation}
b=Bh.  \label{our}
\end{equation}%
Thus, the quantum state-preparation problem can be phrased in a way
similar to the QLSP, i.e.:

\textbf{QSPP}: Given a matrix $B$ and a vector $h$, find an approximation of
the quantum state $\left\vert b\right\rangle $ corresponding to the vector $%
b $.

Comparing Eq. (\ref{our}) with Eq. (\ref{HHL-1}), we can see that with the
mapping%
\begin{eqnarray}
b &\longleftrightarrow &x,  \nonumber \\
B &\longleftrightarrow &A^{-1},  \nonumber \\
h &\longleftrightarrow &b,
\end{eqnarray}%
the original HHL algorithm can be applied here to obtain the approximation
of $\left\vert b\right\rangle $\ from known $B$ and $h$ in $O(poly(\log
(N))) $ runtime, except that the matrix $B$ does not need to be inverted
during the process. Also, preparing the quantum state $\left\vert
h\right\rangle $ corresponding to $h$ requires $n=\log (N)$ single-qubit
quantum gates only (to be shown below). Thus, the total runtime of preparing
$\left\vert b\right\rangle $\ remains at $O(poly(\log (N)))$, so that it
will preserve the exponential speedup for further using the HHL algorithm to
obtain $\left\vert x\right\rangle $.

Note that like the matrix $A$ in the original HHL algorithm, the matrix $B$
is not unitary either. Therefore, it cannot be applied directly on the
quantum state $\left\vert h\right\rangle $ to obtain $\left\vert
b\right\rangle $. Instead, the same treatment on $A$ in \cite{qi2066} is
needed. That is, $B$ can be transformed into $e^{iBt}$ which is unitary so
that it becomes physically implementable. Also, when any of $b_{1}$, ... ,$%
b_{N}$ is a complex number, $B$ is not a Hermitian matrix. Then we need to
apply the trick in Eq. (1) of \cite{qi2066} to expand it as%
\begin{equation}
\tilde{B}=\left[
\begin{array}{cc}
0 & B^{\dagger } \\
B & 0%
\end{array}%
\right]
\end{equation}%
which is Hermitian. Solving%
\begin{equation}
\tilde{b}=\tilde{B}\left[
\begin{array}{c}
h \\
0%
\end{array}%
\right]
\end{equation}%
will obtain%
\begin{equation}
\tilde{b}=\left[
\begin{array}{c}
0 \\
b%
\end{array}%
\right] .
\end{equation}%
For simplicity, in the following we take $B$ as Hermitian to illustrate our
algorithm. In real applications where $B$ is not Hermitian, then apply the
above reduction and the desired $\left\vert b\right\rangle $\ can still be
obtained.

\section{Our state-preparation algorithm}

With the above idea in mind, our algorithm for state-preparation can be
constructed following the HHL algorithm, with two modifications only. Ref.
\cite{ml91} offered a clear and concise step-by-step walkthrough of the
HHL algorithm. Here we use the same notation and workflow as those in
section II of \cite{ml91}, so that it will be easy to compare our algorithm with the original HHL algorithm and see the difference.

As illustrated in Fig.1, our algorithm takes $n_{b}$ target qubits where the final target state $%
\left\vert b\right\rangle $ of the preparation will be stored, $n_{c}$ clock
qubits (as called in \cite{ml91}), and another ancilla qubit. At the
beginning, all these $n_{b}+n_{c}+1$\ qubits are initialized as%
\begin{equation}
\left\vert \Psi _{0}\right\rangle =\left\vert 0...0\right\rangle
_{b}\left\vert 0...0\right\rangle _{c}\left\vert 0\right\rangle _{a}.
\end{equation}%
Then the state-preparation algorithm goes as follows.

%
%


\begin{figure*}[tbp]
\includegraphics[scale=1.25]{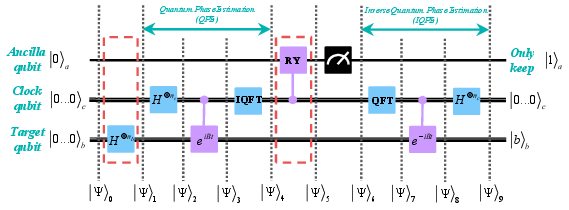}
\caption{Diagram of the quantum circuit of our state-preparation
algorithm, following Fig.1 of \cite{ml91}. Our modifications are marked by the two red dashed boxes.}
\label{fig:epsart}
\end{figure*}


(1) Unlike the original HHL algorithm, our first modification is to apply a
Hadamard gate $H$ on each of the target qubits to transform them into the
quantum state $\left\vert h\right\rangle $ corresponding to $h$ up to a
normalization factor, i.e.,%
\begin{equation}
\left\vert h\right\rangle =\frac{1}{\sqrt{2^{n_{b}}}}\sum%
\limits_{j=0}^{2^{n_{b}}-1}\left\vert j\right\rangle =\frac{1}{\sqrt{%
2^{n_{b}}}}(1,1,...,1,1)^{T}.
\end{equation}%
It takes $n_{b}$ single-qubit gates in total, and the state of the whole
system becomes%
\begin{eqnarray}
\left\vert \Psi _{1}\right\rangle &=&(H^{\otimes n_{b}}\otimes I^{\otimes
n_{c}}\otimes I)\left\vert \Psi _{0}\right\rangle  \nonumber \\
&=&\frac{1}{\sqrt{2^{n_{b}}}}(\left\vert 0\right\rangle +\left\vert
1\right\rangle )^{\otimes n_{b}}\left\vert 0...0\right\rangle _{c}\left\vert
0\right\rangle _{a}  \nonumber \\
&=&\left( \frac{1}{\sqrt{2^{n_{b}}}}\sum\limits_{j=0}^{2^{n_{b}}-1}\left%
\vert j\right\rangle \right) \left\vert 0...0\right\rangle _{c}\left\vert
0\right\rangle _{a}  \nonumber \\
&=&\left\vert h\right\rangle \left\vert 0...0\right\rangle _{c}\left\vert
0\right\rangle _{a}.
\end{eqnarray}

(2) From now on, we follow the HHL algorithm except where noted. That is, we
launch the quantum phase estimation (QPE) process, which starts from
applying a Hadamard gate $H$ on each of the clock qubits, turning the state
into%
\begin{eqnarray}
\left\vert \Psi _{2}\right\rangle &=&(I^{\otimes n_{b}}\otimes H^{\otimes
n_{c}}\otimes I)\left\vert \Psi _{1}\right\rangle  \nonumber \\
&=&\left\vert h\right\rangle \left( \frac{1}{\sqrt{2^{n_{c}}}}%
\sum\limits_{k=0}^{2^{n_{c}}-1}\left\vert k\right\rangle \right) \left\vert
0\right\rangle _{a}.
\end{eqnarray}

(3) The controlled rotation part of the QPE: apply $n_{c}$\ controlled gates
on $\left\vert h\right\rangle $, as shown in Fig. 2 of \cite{ml91}. That is,
the $r$th controlled gate ($r=n_{c}-1,n_{c}-2,...,0$) performs the
transformation $U^{2^{r}}$ on $\left\vert h\right\rangle $, with the $r$th
clock qubit as the control qubit. Here the unitary transformation $U$ is
defined as%
\begin{equation}
U=e^{iBt}.
\end{equation}%
The state becomes%
\begin{eqnarray}
\left\vert \Psi _{3}\right\rangle  &=&(QPE\otimes I)\left\vert \Psi
_{2}\right\rangle   \nonumber \\
&=&\sum\limits_{j=0}^{2^{n_{b}}-1}\frac{1}{\sqrt{2^{n_{b}}}}\left\vert
j\right\rangle   \nonumber \\
&&\otimes \left( \frac{1}{\sqrt{2^{n_{c}}}}\sum%
\limits_{k=0}^{2^{n_{c}}-1}e^{2\pi i\phi _{j}k}\left\vert k\right\rangle
\right) \left\vert 0\right\rangle _{a},  \label{QPE}
\end{eqnarray}%
where%
\begin{equation}
\phi _{j}=\frac{b_{j}t}{2\pi }
\end{equation}%
with $t$ being the parameter in the HHL algorithm. According to \cite{qi2066}%
, the operator $e^{iBt}$ can be simulated using the approach in \cite{qi2469}%
\ in time $\tilde{O}[\log (N)s^{2}t]$ (where the $\tilde{O}$\ suppresses
more slowly growing terms) when the matrix $B$ is $s$ sparse. In our case,
from Eq. (\ref{B}) we know that $s=1$. (To understand how Eq. (\ref{QPE}) is
obtained, it is highly recommended to read the derivation of Eqs. (14) and
(20) of \cite{ml91}, and note the fact that $b_{j}$ ($j=1,...,N$) are
exactly the eigenvalues of our matrix $B$ in the computational basis, and $%
\left\vert h\right\rangle $ is a uniform superposition of the eigenvectors.)

(4) Perform the inverse quantum Fourier transform (IQFT) on the clock
qubits. This ends the QPE and the resultant state (as analog to Eq. (21) of
\cite{ml91}) is%
\begin{eqnarray}
\left\vert \Psi _{4}\right\rangle &=&(I^{\otimes n_{b}}\otimes IQFT\otimes
I)\left\vert \Psi _{3}\right\rangle  \nonumber \\
&=&\sum\limits_{j=0}^{2^{n_{b}}-1}\frac{1}{\sqrt{2^{n_{b}}}}\left\vert
j\right\rangle \left\vert \tilde{\lambda}_{j}\right\rangle \left\vert
0\right\rangle _{a}
\end{eqnarray}%
with%
\begin{equation}
\tilde{\lambda}_{j}=N\phi _{j}=N\frac{b_{j}t}{2\pi }.
\end{equation}

(5) Rotate the ancilla qubit using the clock qubits as control qubits. In
the original HHL algorithm, this rotation transforms $\left\vert \Psi
_{4}\right\rangle $ into%
\begin{equation}
\left\vert \Psi _{5}^{\prime }\right\rangle =\sum\limits_{j=0}^{2^{n_{b}}-1}%
\frac{1}{\sqrt{2^{n_{b}}}}\left\vert j\right\rangle \left\vert \tilde{\lambda%
}_{j}\right\rangle (\sqrt{1-\frac{C^{2}}{\tilde{\lambda}_{j}^{2}}}\left\vert
0\right\rangle _{a}+\frac{C}{\tilde{\lambda}_{j}}\left\vert 1\right\rangle
_{a}).  \label{origin}
\end{equation}%
But here we introduce our second and final modification to the HHL algorithm.
Instead of Eq. (\ref{origin}), we replace the controlled rotation with
another one which results in%
\begin{equation}
\left\vert \Psi _{5}\right\rangle =\sum\limits_{j=0}^{2^{n_{b}}-1}\frac{1}{%
\sqrt{2^{n_{b}}}}\left\vert j\right\rangle \left\vert \tilde{\lambda}%
_{j}\right\rangle (\sqrt{1-C^{2}\tilde{\lambda}_{j}^{2}}\left\vert
0\right\rangle _{a}+C\tilde{\lambda}_{j}\left\vert 1\right\rangle _{a}).
\label{step5}
\end{equation}%
Here the real constant $C$ should be chosen to ensure that $\left\vert C%
\tilde{\lambda}_{j}\right\vert ^{2}\leq 1$\ for all $j=0,...,2^{n_{b}}-1$.
According to Sec. 4.3 of \cite{qi2455}, this rotation can be implemented
using $O[(\log \epsilon _{1}^{-1})^{4}]$ quantum operations and $O[(\log
\epsilon _{1}^{-1})^{3}]$ qubits with error $\epsilon _{0}=\epsilon _{1}^{2}$%
. In other words, when the desired error is $\epsilon _{0}$, we need $%
n_{c}=O[(\log \epsilon _{0}^{-1/2})^{3}]$ clock qubits and $O[(\log \epsilon
_{0}^{-1/2})^{4}]=O(n_{c}^{4/3})$ quantum operations.%

(6) Now return to the HHL algorithm to disentangle the target qubits from
other qubits. Namely, we first measure the ancilla qubit in the basis $%
\{\left\vert 0\right\rangle _{a},\left\vert 1\right\rangle _{a}\}$. If the
result is $\left\vert 0\right\rangle _{a}$ then the state-preparation fails
and we need to restart the whole process over again. Else if the result is $%
\left\vert 1\right\rangle _{a}$\ then we continue. The state of the system
in this case is%
\begin{eqnarray}
\left\vert \Psi _{6}\right\rangle &=&\frac{1}{\sqrt{\sum%
\limits_{j=0}^{2^{n_{b}}-1}\left\vert \frac{C\tilde{\lambda}_{j}}{\sqrt{%
2^{n_{b}}}}\right\vert ^{2}}}\sum\limits_{j=0}^{2^{n_{b}}-1}\frac{C\tilde{%
\lambda}_{j}}{\sqrt{2^{n_{b}}}}\left\vert j\right\rangle \left\vert \tilde{%
\lambda}_{j}\right\rangle \left\vert 1\right\rangle _{a}  \nonumber \\
&=&\frac{1}{\sqrt{\sum\limits_{j=0}^{2^{n_{b}}-1}\left\vert \tilde{\lambda}%
_{j}\right\vert ^{2}}}\sum\limits_{j=0}^{2^{n_{b}}-1}\tilde{\lambda}%
_{j}\left\vert j\right\rangle \left\vert \tilde{\lambda}_{j}\right\rangle
\left\vert 1\right\rangle _{a}.
\end{eqnarray}

(7) Perform the inverse quantum phase estimation (IQPE) process, starting by
applying the quantum Fourier transform (QFT) on the clock qubits. The
resultant state (as analog to Eq. (24) of \cite{ml91}) is%
\begin{eqnarray}
\left\vert \Psi _{7}\right\rangle  &=&(I^{\otimes n_{b}}\otimes QFT\otimes
I)\left\vert \Psi _{6}\right\rangle   \nonumber \\
&=&\frac{1}{\sqrt{\sum\limits_{j=0}^{2^{n_{b}}-1}\left\vert \tilde{\lambda}%
_{j}\right\vert ^{2}}}\sum\limits_{j=0}^{2^{n_{b}}-1}\tilde{\lambda}%
_{j}\left\vert j\right\rangle QFT\left\vert \tilde{\lambda}_{j}\right\rangle
\left\vert 1\right\rangle _{a}  \nonumber \\
&=&\frac{1}{\sqrt{\sum\limits_{j=0}^{2^{n_{b}}-1}\left\vert \tilde{\lambda}%
_{j}\right\vert ^{2}}}\sum\limits_{j=0}^{2^{n_{b}}-1}\tilde{\lambda}%
_{j}\left\vert j\right\rangle   \nonumber \\
&&\otimes \left( \frac{1}{2^{n_{c}/2}}\sum_{y=0}^{2^{n_{c}}-1}e^{2\pi iy%
\tilde{\lambda}_{j}/N}\left\vert y\right\rangle \right) \left\vert
1\right\rangle _{a}.
\end{eqnarray}

(8) The controlled rotation part of the IQPE: similar to step (3), $n_{c}$\
controlled gates are applied on the target qubits with the clock qubits as
the control qubits. But the unitary transformation $U$ is replaced by $%
U^{-1}=e^{-iBt}$. The result is%
\begin{eqnarray}
\left\vert \Psi _{8}\right\rangle  &=&(IQPE\otimes I)\left\vert \Psi
_{7}\right\rangle   \nonumber \\
&=&\frac{1}{2^{n_{c}/2}\sqrt{\sum\limits_{j=0}^{2^{n_{b}}-1}\left\vert
\tilde{\lambda}_{j}\right\vert ^{2}}}\sum\limits_{j=0}^{2^{n_{b}}-1}\tilde{%
\lambda}_{j}\left\vert j\right\rangle   \nonumber \\
&&\otimes \left( \sum_{y=0}^{2^{n_{c}}-1}e^{-ib_{j}ty}e^{2\pi iy\tilde{%
\lambda}_{j}/N}\left\vert y\right\rangle \right) \left\vert 1\right\rangle
_{a} \\
&=&\frac{1}{2^{n_{c}/2}\sqrt{\sum\limits_{j=0}^{2^{n_{b}}-1}\left\vert
b_{j}\right\vert ^{2}}}\sum\limits_{j=0}^{2^{n_{b}}-1}b_{j}\left\vert
j\right\rangle (\sum_{y=0}^{2^{n_{c}}-1}\left\vert y\right\rangle
)\left\vert 1\right\rangle _{a}  \nonumber \\
&=&\sum\limits_{j=0}^{2^{n_{b}}-1}b_{j}\left\vert j\right\rangle (\frac{1}{%
2^{n_{c}/2}}\sum_{y=0}^{2^{n_{c}}-1}\left\vert y\right\rangle )\left\vert
1\right\rangle _{a}.
\end{eqnarray}%
where we made use of $\tilde{\lambda}_{j}=Nb_{j}t/2\pi $ and $%
\sum\nolimits_{j=0}^{2^{n_{b}}-1}\left\vert b_{j}\right\vert ^{2}=1$.

(9) End the IQPE by applying a Hadamard gate $H$ on each of the clock
qubits. This completes the whole preparation process and we eventually yield%
\begin{eqnarray}
\left\vert \Psi _{9}\right\rangle  &=&(I^{\otimes n_{b}}\otimes H^{\otimes
n_{c}}\otimes I)\left\vert \Psi _{8}\right\rangle   \nonumber \\
&=&\left( \sum\limits_{j=0}^{2^{n_{b}}-1}b_{j}\left\vert j\right\rangle
\right) \left\vert 0...0\right\rangle _{c}\left\vert 1\right\rangle _{a}
\nonumber \\
&=&\left\vert b\right\rangle \left\vert 0...0\right\rangle _{c}\left\vert
1\right\rangle _{a}.
\end{eqnarray}%
That is, the target state $\left\vert b\right\rangle $ of the preparation is
successfully stored in the target qubits.

\bigskip
\bigskip

\section{Error and runtime}

As shown in step (5), when the desired error of the quantum operation for
achieving Eq. (\ref{step5}) is $\epsilon _{0}$, it takes $n_{c}=O[(\log
\epsilon _{0}^{-1/2})^{3}]$ clock qubits and $O[(\log \epsilon
_{0}^{-1/2})^{4}]=O(n_{c}^{4/3})$ quantum operations.
The error of other steps is the same as that of the original HHL algorithm,
which was analyzed in \cite{qi2066}. That is, the dominant source of error
is phase estimation, which can be done with error $\epsilon $ in time
proportional to%
\begin{equation}
ts^{2}(t/\epsilon )^{o(1)}=:\tilde{O}(ts^{2})
\end{equation}%
for a $s$ sparse matrix $A$.

Therefore, to reach a successful computation, the runtime of the original
HHL algorithm is \cite{qi2066}%
\begin{equation}
\tilde{O}(\log (N)s^{2}\kappa ^{2}/\epsilon ).
\end{equation}%
In our case, $A$ is replaced by $B$ and we have $s=1$ in the above two
equations, as can be seen from Eq. (\ref{B}). Meanwhile, our first
modification (step (1)) takes $n_{b}$ Hadamard gates. The second
modification (step (5)) replaces Eq. (\ref{origin}) with Eq. (\ref{step5}),
which takes $O(n_{c}^{4/3})$ quantum operations to implement. Putting these
three parts together, the total runtime of our state-preparation algorithm
remains at $O(\log (N))$, as long as the condition number $\kappa $ of the
matrix $B$ does not grow like $N^{c}$ (in the case where $\kappa $\ is
large, the treatment in \cite{qi2466} may help). Therefore, for preparing
quantum states satisfying this condition, exponential speedup can be achieved
in comparison with other existing state-preparation methods.

\section{Discussion}

In conclusion, our modified algorithm can prepare a quantum state with a
runtime at the same level of the original HHL algorithm. Thus, using it to
prepare the initial state will preserve the exponential speedup advantage of
the HHL algorithm.

Note that among the four caveats that we mentioned at the beginning of this
paper, our approach conquers the first one only. To ensure that the speedup
brought by the HHL algorithm will not be ruined, the other three caveats
still need to be satisfied. Nevertheless, as our approach solves the
encoding bottleneck, it surely extends the potential application range of
the HHL algorithm. Also, in other application scenarios which have intensive
needs of preparing initial states, e.g., quantum machine learning, our
approach can serve as a standalone solution to speed up the process.

\section*{Acknowledgements}

This work was supported in part by Guangdong Basic and Applied Basic
Research Foundation (Grant No. 2019A1515011048).

%
%
%
%
%
%
%
%
%
%
%
%
%
%

\end{document}